\documentclass[11pt,conference]{IEEEconf}

\input epsf
\usepackage{graphicx}

\usepackage{cite}
\usepackage{amsmath,amssymb,amsfonts}
\usepackage[hidelinks]{hyperref}
\usepackage{array}
\usepackage{algorithmic}
\usepackage{graphicx}
\usepackage{textcomp}
\usepackage[T1]{fontenc}
\usepackage{tgbonum}
\usepackage{xcolor}
\usepackage{caption}
\usepackage{subcaption}
\usepackage{multirow, multicol}
\usepackage{array,multirow,graphicx}
 \usepackage{float}
\begin{document}


\title{\Large Harnessing the Power of LLMs in Source Code Vulnerability Detection}

\author{Andrew Arash Mahyari,~\textit{Member, IEEE}\\
	\normalsize AIVault Inc. \\
	\normalsize andrew.mahyari@ai-vault.com
}


\maketitle
\begin{abstract}
Software vulnerabilities, caused by unintentional flaws in source code, are a primary root cause of cyberattacks. Static analysis of source code has been widely used to detect these unintentional defects introduced by software developers. Large Language Models (LLMs) have demonstrated human-like conversational abilities due to their capacity to capture complex patterns in sequential data, such as natural languages. In this paper, we harness LLMs' capabilities to analyze source code and detect known vulnerabilities. To ensure the proposed vulnerability detection method is universal across multiple programming languages, we convert source code to LLVM IR and train LLMs on these intermediate representations. We conduct extensive experiments on various LLM architectures and compare their accuracy. Our comprehensive experiments on real-world and synthetic codes from NVD and SARD demonstrate high accuracy in identifying source code vulnerabilities. 

\end{abstract}
\IEEEoverridecommandlockouts

%
\IEEEpeerreviewmaketitle



\section{Introduction}


When developing source code, software developers often overlook flaws that can lead to vulnerabilities. These vulnerabilities open the door for cyber attacks on software, systems, and applications, potentially causing severe societal and financial consequences \cite{arnold2017after, morrison2018vulnerabilities}. Each year, numerous vulnerabilities are documented in the Common Vulnerabilities and Exposures (CVE) database \cite{cve}.

\begin{figure*}[t]
    \centering
    \includegraphics[width=0.8\textwidth]{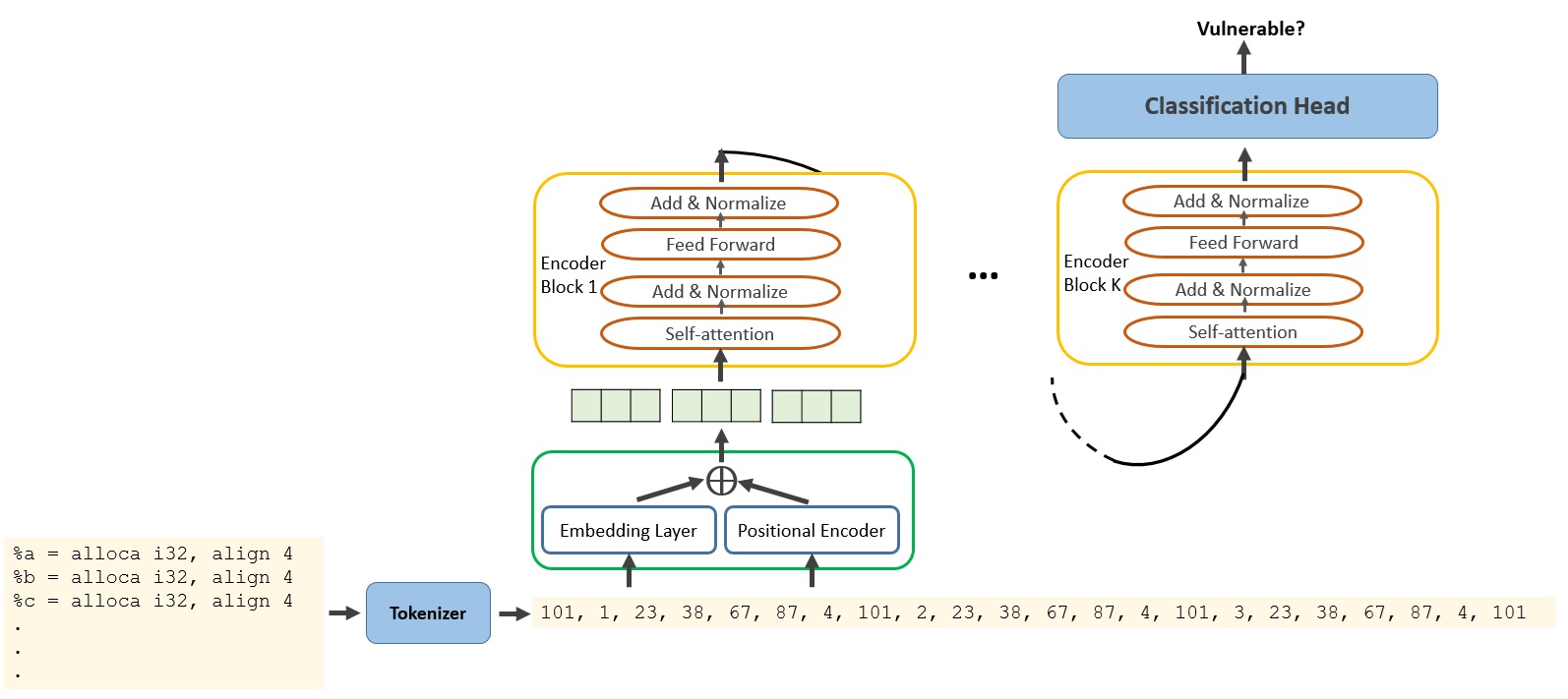}
    \caption{The overall architecture of the proposed vulnerability detection algorithm. First, source codes are converted to LLVM IRs, then LLVM IRs are converted to iSeVCs, and then the tokenizer converts them to unique IDs. The output of LLM is used to predict whether the whole source code is vulnerable.}
    \label{fig:overal}
    \vspace{-2mm}
\end{figure*}

Source code-based static analysis has been employed to detect vulnerabilities, with methods falling into several categories such as code similarity-based approaches \cite{kim2017vuddy, li2016vulpecker, du2019leopard, wanwarang2020testing} and pattern-based methods \cite{grieco2016toward, yamaguchi2013chucky, yamaguchi2012generalized}. Machine learning, including deep learning approaches, also falls under the pattern-based category \cite{li2021vuldeelocator, li2021sysevr, sedaghatbaf2021automated}. These approaches identify vulnerability patterns in source code and can generalize these patterns to detect previously unseen vulnerabilities. Inspired by the success of deep learning in computer vision, some studies have applied it to source code vulnerability detection \cite{li2021vuldeelocator, li2021sysevr, li2018vuldeepecker}. Recent studies have leveraged LLMs to detect vulnerabilities in source codes \cite{li2024llm, lu2024grace}. VulDeePecker \cite{li2018vuldeepecker} was the pioneering approach to utilize deep learning for vulnerability detection, focusing solely on data dependencies among source code lines, which limited its ability to achieve fine-grained detection. To address these limitations, SySeVR (Syntax-based, Semantics-based, and Vector Representations) \cite{li2021sysevr} introduced semantic-based vulnerability candidates and vector representations of source code, representing the first deep learning application for vulnerability detection. Subsequently, VulDeeLocator \cite{li2021vuldeelocator} was developed to enhance VulDeePecker by enabling finer-grained detection, pinpointing specific code segments responsible for vulnerabilities.

There are several shortcomings in previous studies that motivate this paper. In earlier research \cite{li2021vuldeelocator}, Lower Level Virtual Machine (LLVM) intermediate representation (IR) lines of code are treated sequentially, like sentences in a text. However, dependencies between lines in source code are not necessarily sequential. For example, the first two lines may define variables as {\fontfamily{qcr}\selectfont int a; int b;}, and the third line might assign a value to the variable defined in the first line {\fontfamily{qcr}\selectfont a=6;}. Here, the third line is related to the first line. This issue becomes even more complex with user-defined functions. Additionally, existing methods cannot precisely identify the specific lines of code that contribute to a vulnerability; they can only identify regions consisting of several lines \cite{li2021vuldeelocator}.

This paper leverages Large Language Models (LLMs) to analyze source code and identify vulnerabilities. Each programming language, like natural language, has its own vocabulary and rules. These vocabularies consist of the commands used in the programming language, and the rules are similar to grammar, explaining how the commands (vocabulary) can be utilized in source code. To ensure our proposed method is universal, we convert source code to LLVM IRs. We follow the experimental setup in \cite{li2021vuldeelocator} to perform this conversion. We then develop a custom tokenizer based on the LLVM IR vocabulary to convert source code into unique identification numbers before feeding them into the LLM. The LLM, equipped with a classification head, is trained end-to-end to detect vulnerabilities. Figure \ref{fig} illustrates the overall architecture of the proposed approach.

\section{Data}
\label{sec:data}


We utilized the dataset collected and processed in \cite{li2021vuldeelocator}. This dataset comprises source codes of C programs from two primary sources of vulnerabilities: NVD \cite{NVD} and SARD \cite{SARD}. Compatible source codes from these sources were compiled into LLVM intermediate representations \cite{llvm}. The dataset includes 14,511 programs, consisting of 2,182 real-world programs and 12,329 synthetic and academic programs from SARD. The real-world programs are open-source C codes, while the synthetic and academic programs are test cases from SARD. The training set includes real-world vulnerable programs reported before 2017, and the test set contains vulnerable codes reported between 2017 and 2019 (vulnerabilities unknown to the training set) \cite{li2021vuldeelocator}. The dataset preparation and conversion for our proposed method involved two key steps, detailed in \cite{li2021vuldeelocator}.

In the first step, Source code-based Syntax Vulnerability Candidates (sSyVCs) are extracted from the source codes. sSyVCs are defined as code segments exhibiting certain vulnerability syntax characteristics. In the second step, Intermediate code-based Semantics Vulnerability Candidates (iSeVCs) are generated from the intermediate codes based on the identified sSyVCs \cite{li2021vuldeelocator}. To extract sSyVCs, the syntax characteristics of known vulnerabilities are represented using abstract syntax trees of the source code. Four types of vulnerability syntax characteristics are considered: Library/API Function Call (FC), Array Definition (AD), Pointer Definition (PD), and Arithmetic Expression (AE) \cite{li2021vuldeelocator}.

In the second step, the Clang compiler is used to generate LLVM bitcode files, link them according to their dependencies, and produce the linked IR files. Given an sSyVC, its dependency graph is created from the linked IR file and then sliced according to the sSyVC (for more details, see \cite{li2021vuldeelocator}). Each local variable is converted to a numeric value with a "\%" prefix. For each function $f_\gamma$ called by function $f_\alpha$, the IR slice of function $f_\gamma$ is appended to its call in function $f_\alpha$, with variable names adjusted accordingly. The resulting IRs, known as iSeVCs, are used for vulnerability detection \footnote{For more information, interested readers may refer to \cite{li2021vuldeelocator}.}.


\subsection{Preprocessing}
\label{sec:pre}

The performance of sSeVC and iSeVC was compared in \cite{li2021vuldeelocator}, and it was demonstrated that iSeVC provides better accuracy. Consequently, we use iSeVC from \cite{li2021vuldeelocator} as the language input in this paper. We remove all user-defined functions (identified by \verb|"call"| lines immediately followed by \verb|"define"| lines in the processed dataset, while retaining the lines within these functions. This approach reduces the number of user-defined vocabulary terms and enhances our algorithm's robustness regarding function names. To expedite training and evaluation, we only include programs with fewer than 265 lines of LLVM IRs. This does not compromise the scalability of the algorithm.

If the source code is marked as vulnerable in \cite{li2021vuldeelocator}, we assign it a label of 1; otherwise, it receives a label of 0. Additionally, the original data from \cite{li2021vuldeelocator} includes the specific line numbers where vulnerabilities occur. For each line of LLVM IR, we create a separate binary label: if the line number corresponds to the vulnerability in the data, its label is set to 1; otherwise, it is set to 0.

\section{Proposed Vulnerability Detector}

The goal of this paper is to detect source code vulnerabilities based on their LLVM IRs. In this section, we describe different sections of the proposed method.

\subsection{Tokenizer}
\label{sec:tokenizer}

LLMs employ tokenizers to preprocess text data before feeding it into the model. Tokenization involves dividing the input text into smaller units known as tokens, which can be words, subwords, or even individual characters, depending on the tokenization strategy. Each token is then mapped to a unique identifier (ID) using a vocabulary that the model has been trained on, allowing the model to process numerical data instead of raw text. Additionally, tokenizers handle special tokens used by the model, such as {\sffamily{[CLS]}} (classification), {\sffamily{[SEP]}} (separator), {\sffamily{[PAD]}} (padding), and others that mark the beginning or end of a sequence or pad sequences to a fixed length. However, in this paper, we use LLMs for source code analysis, and standard tokenizers are not suitable for converting source codes to IDs.

We create a vocabulary from all tokens separated by a single space, resulting in a total of $20,086$ unique tokens. This indicates that there are only $20,086$ different tokens across all source codes converted to LLVM IR. Each line of code is treated as a sentence and is marked with special tokens {\sffamily{[SEP]}} at the beginning and end. Each token is assigned a unique ID, resulting in 20,086 unique IDs for the tokens, and additional unique IDs for the special tokens. The source code is then converted into a sequence of these unique IDs, separated by the ID of the {\sffamily{[SEP]}} token. This sequence of IDs is used as input to the LLM.

Let ${\bf X}_i={ x_i(1), x_i(2), \ldots, x_i(L_i) }$ represent the sequence of unique IDs for the tokens in the $i$th piece of code, where $x_i(j)$ is the unique ID of the $j$th token, and $L_i$ is the total number of tokens in this code. Additionally, let $Y_i \in {0, 1}$ denote the label for the $i$th piece of code, where $1$ indicates that the code is vulnerable and $0$ indicates no vulnerability. Furthermore, let ${ y_i(1), y_i(2), \ldots, y_i(t) }$ represent the labels for each line of code, where $y_i(\cdot) \in {0, 1}$.

\subsection{Large Language Model}

LLMs have proven highly effective in natural language processing tasks due to their ability to uncover complex relationships between words in text sequences. By leveraging the transformer's capability to capture contextual patterns, we can learn the patterns of safe source code and differentiate them from vulnerable code. This process is driven by the self-attention mechanism \cite{Attention}, which allows each token in the input sequence to reference and weigh other tokens within that sequence when generating its output representation. This mechanism assigns a significance score to each token, indicating the degree of attention it should receive. The model architecture can be divided into the following main components:

\noindent \textbf{Embedding Layer:} The embedding layer serves as the initial conversion phase between the input data and the transformer. The input sequence of unique IDs, ${\bf X}_i$, is transformed into a sequence of vector embedding $\mathbf{E}_i$, where each element is $d$ dimensional. Let $f_e(\cdot)$ represent the embedding layer which converts unique IDs into vector representations, then $\mathbf{E}_i=f_e({\bf X}_i)$.

\noindent \textbf{Positional Encoder:} A positional encoder is a mechanism that provides positional information for each token in a sequence. Unlike RNNs and CNNs, transformers lack an inherent sequential structure, so they rely on positional encodings to understand token order. This encoding allows the model to differentiate between tokens based on their positions, which is crucial for tasks where word order impacts meaning, such as in source code analysis. Positional encodings are added to the input token embeddings before feeding them into the transformer layers. This combined input gives the model both the content (from the token embeddings) and the positional information (from the encodings), enabling effective sequence processing.
Positional encodings are generated using sinusoidal functions, which provide a fixed encoding for each position. These encodings are calculated using sine and cosine functions of different frequencies, allowing the model to generalize to sequences of varying lengths since the encoding is not learned and can be applied to any position. The positional encoding vector is designed to have the same dimension, $d$, as the embedding layer, allowing them to be summed together.

\noindent \textbf{Encoder:} After the positional encoding and token embedding are combined, they are passed to the encoder. The encoder is composed of several blocks, each containing a feedforward network and a multi-head self-attention mechanism. The feedforward network is a standard fully connected neural network. The self-attention mechanism, however, distinguishes transformers from other traditional deep learning architectures. It allows the model to capture dependencies and relationships across the entire input sequence, regardless of the distance between tokens. By weighing the importance of different tokens based on both their content and their positions within the sequence, the self-attention mechanism becomes highly effective for sequential tasks. This process is achieved through the following steps:

\textbf{\textit{1- Input Linear Transformation:}} Once the input sequence has been converted into embeddings, each vector \( \mathbf{e}(t) \) is transformed into three distinct vectors: queries (\( \mathbf{Q} \)), keys (\( \mathbf{K} \)), and values (\( \mathbf{V} \)). This transformation is accomplished using the weight matrices \( \mathbf{W}_Q \), \( \mathbf{W}_K \), and \( \mathbf{W}_V \), which are learned during the training process. These transformations are fundamental to the self-attention mechanism of the transformer model and can be described by the following equations:

\begin{equation}
    \mathbf{Q} = \mathbf{W}_Q\mathbf{E}, \hspace{1mm} \mathbf{K} = \mathbf{W}_K\mathbf{E}, \hspace{1mm} \mathbf{V} = \mathbf{W}_V\mathbf{E}
\end{equation}


\noindent where \( \mathbf{E} \) is the matrix of embedded vectors, and \( \mathbf{W}_Q \), \( \mathbf{W}_K \), and \( \mathbf{W}_V \) are the weight matrices that transform the embedded vectors into queries, keys, and values, respectively.

\textbf{\textit{2- Computation of Attention Scores:}} Attention scores are calculated by taking the dot product of the query vector with each key vector, and then scaling by \( \frac{1}{\sqrt{d_k}} \), where \( d_k \) is the dimension of the key vectors. The softmax function is then applied to these scores to obtain a probability distribution.

\vspace{-3mm}

\begin{align}
\text{Attention}(\mathbf{Q}, \mathbf{K}, \mathbf{V}) &= \text{softmax}\left(\frac{\mathbf{Q}\mathbf{K}^T}{\sqrt{d_k}}\right)\mathbf{V}
\end{align}

\vspace{-1mm}

The final output aggregates the input features, highlighting the most relevant information for each token in the sequence. This mechanism enables the model to capture both a global understanding and the intricate details needed to comprehend the input patterns.

\noindent \textbf{Classification Head:} Since the goal of this algorithm is to create a classifier using LLMs, we employ a classification head on top of the encoder. The final hidden state of the encoder is fed into the classification head, which predicts whether the code is vulnerable or not. The classification head may consist of one or several linear layers with activation functions. This layer learns and predict $p(Y_i | X_i)$. The whole architecture is trained end-to-end. The loss function is \textit{cross entropy} with the stochastic gradient descent optimization algorithm \cite{Goodfellow-et-al-2016}.

\begin{table}[t]
\centering
\caption{The accuracy of different LLM architectures with the state-of-the-art source code vulnerability detectors on the test dataset.}
\label{tab:whole-all-target}
\begin{tabular}{ cc } 
\hline
Methods & Accuracy \\
\hline
Bert & 98.25 \\
DistilBert & 98.17 \\
LSTM \cite{mahyari2022hierarchical} & 98.10 \\
VulDeeLocator \cite{li2021vuldeelocator} & 82.43 \\
\hline
\vspace{-6mm}
\end{tabular}
\end{table}


\begin{figure*}
     \centering
     \begin{subfigure}[b]{0.3\textwidth}
         \centering
         \includegraphics[width=\textwidth]{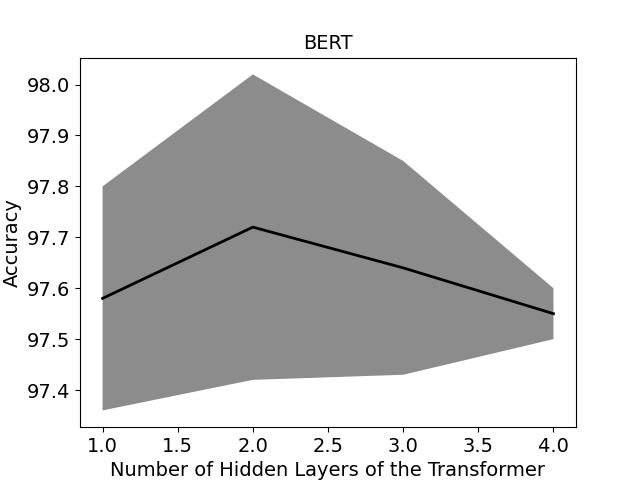}
         \caption{}
         \label{fig:bert}
     \end{subfigure}
     \begin{subfigure}[b]{0.3\textwidth}
         \centering
         \includegraphics[width=\textwidth]{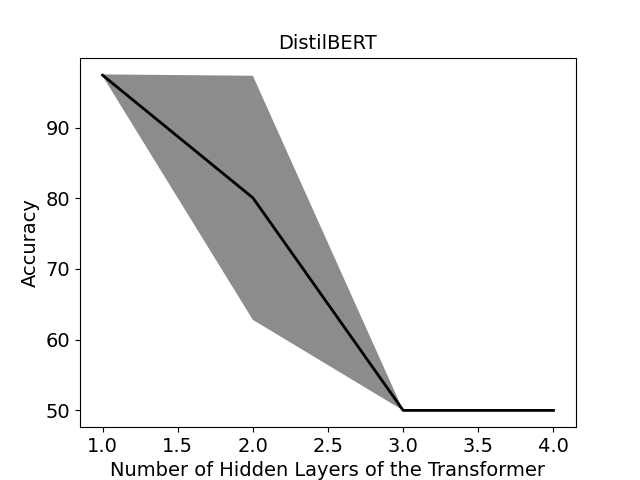}
         \caption{}
         \label{fig:distl}
     \end{subfigure}
     \hfill
        \caption{The accuracy vs number of FC layers of the classifier head: (a) Bert; (b) DistilBert.}
        \label{fig:ablation}
\end{figure*}

\section{Experimental Results}
\label{sec:experiment}

In this section, we assess the effectiveness of the proposed approach in detecting source code vulnerabilities. The dataset is compiled in the LLVM environment to obtain their LLVM intermediate representation, which is then converted to iSeVCs as detailed in Section~\ref{sec:data}. The training dataset contains significantly fewer vulnerable samples compared to benign ones. During the sampling process, we randomly and equally sample from both vulnerable and benign samples to create the training subset. The trained model is then evaluated on all samples in the test dataset. This experiment is also conducted using the same dataset with two other methods VulDeeLocator \cite{li2021vuldeelocator} and an LSTM-based method \cite{mahyari2022hierarchical} for comparison.

We utilized two LLM architectures: Bert \cite{devlin2018bert} and DistilBert \cite{sanh2019disilbert}, and compared their performance with two state-of-the-art methods. Classifier heads composed of fully connected layers were added on top of the LLMs to map the vector representations provided by the LLMs onto the label space. Table~\ref{tab:whole-all-target} shows the accuracy of the four methods on the test dataset. The results indicate that NLP-based methods achieve high accuracy. The accuracy of the Bert and DistilBert methods is comparable to that of the LSTM-based method, whereas VulDeeLocator exhibits significantly lower accuracy. The superior accuracy of NLP-based methods can be attributed to their ability to capture the complex structure of symbolic data sequences (i.e., natural language), making them well-suited for capturing the structure of source codes. Additionally, the NLP-based approaches are trained end-to-end, whereas VulDeeLocator \cite{li2021vuldeelocator} employs a two-step training process: first training a word2vec model separately from the classifier module, and then training a classifier to detect vulnerable source codes.

\vspace{-2mm}

\subsection{Ablation Study}

Here we conduct an in-depth ablation study on the number of layers in the classifier head of LLMs. The objective of this study is to determine the impact of the classifier head's depth on the accuracy of the outcome. We used the same LLVM IRs described earlier for this experiment. At each iteration, one fully connected (FC) layer with a ReLU activation function is added to the classifier head, and the model is then trained and evaluated on the LLVM IRs. The experiment is repeated five times for each number of FC layers. Figure~\ref{fig:ablation} shows the mean and one standard deviation of the accuracy for different numbers of FC layers. As shown in Figure~\ref{fig:ablation}.a, the classifier head with two FC layers yields the highest accuracy for Bert. However, the accuracy does not significantly drop as the number of layers increases. In contrast, the number of FC layers plays a crucial role in the accuracy of the DistilBert classifier. As the number of layers increases, the accuracy decreases, with the best results achieved using only one FC layer in the classifier head.

\vspace{-2mm}

\section{Conclusion}
\vspace{-1mm}

Large language models (LLMs) have opened new possibilities for analyzing and generating source codes, enabling their use in detecting vulnerabilities, recommending next commands, or converting programming languages. In this paper, we proposed a vulnerability detection method based on LLMs to represent source codes. These representations were used to train LLMs to identify vulnerabilities within source codes. The proposed architecture achieved significant accuracy, attributable to the vocabulary built from LLVM IR source codes. Although this paper examines the hyperparameters of the classifier head and various LLM architectures, further comprehensive studies are needed to explore the generalizability of LLMs in detecting all types of vulnerabilities.



%


\begin{thebibliography}{1}
\vspace{-2mm}
\bibitem {mahyari2022hierarchical}
A. Mahyari, "A Hierarchical Deep Neural Network for Detecting Lines of Codes with Vulnerabilities," 22nd IEEE International Conference on Software Quality, Reliability, and Security (QRS 2022), Dec 2022.

\bibitem {li2021vuldeelocator}
Li, Z., and Zou, D., and Xu, S., and Chen, Z., and Zhu, Y., and Jin, H. ``Vuldeelocator: a deep learning-based fine-grained vulnerability detector,'' IEEE Transactions on Dependable and Secure Computing,'' vol. 19, no. 14, pp. 2821 - 2837, 2022

\bibitem{li2024llm}
Z. Li, S. Dutta, M. Naik, "LLM-Assisted Static Analysis for Detecting Security Vulnerabilities", arXiv preprint arXiv:2405.17238, 2024.

\bibitem{lu2024grace}
G. Lu, X. Ju, X. Chen, W. Pei, Z. Cai, "GRACE: Empowering LLM-based software vulnerability detection with graph structure and in-context learning", Journal of Systems and Software, 212, 112031, 2024.


\bibitem{arnold2017after}
C. Arnold, “After equifax hack, calls for big changes in credit reporting industry,” 2017.

\bibitem{morrison2018vulnerabilities}
P. J Morrison, and R. Pandita, and X. Xiao, and R. Chillarege, and L. Williams, “Are vulnerabilities discovered and resolved like other defects?,”
Empirical Software Engineering, vol. 23(3), 2018.

\bibitem{cve}
``Cve Common vulnerabilities and exposures,'' 2018.

\bibitem{kim2017vuddy}
S. Kim, and S. Woo, and H. Lee, and H. Oh, “Vuddy: A scalable approach for vulnerable code clone discovery,” in 2017 IEEE Symposium
on Security and Privacy (SP). IEEE, 2017, pp. 595–614.

\bibitem{li2016vulpecker}
Z. Li, and D. Zou, and S. Xu, and H. Jin, H. Qi, and J. Hu, “Vulpecker: an automated vulnerability detection system based on code similarity
analysis,” in Proceedings of the 32nd Annual Conference on Computer Security Applications, 2016, pp. 201–213.

\bibitem{du2019leopard}
X. Du, and B. Chen, and Y. Li, and J. Guo, and Y. Zhou, and Y. Liu, and Y. Jiang, “Leopard: Identifying vulnerable code for vulnerability assessment
through program metrics,” in 2019 IEEE/ACM 41st International Conference on Software Engineering (ICSE). IEEE, 2019, pp.
60–71.

\bibitem{wanwarang2020testing}
T. Wanwarang, and N. P Borges Jr, and L. Bettscheider, and A. Zeller, “Testing apps with real-world inputs,” in Proceedings of the
IEEE/ACM 1st International Conference on Automation of Software Test, 2020, pp. 1–10.

\bibitem{grieco2016toward}
G. Grieco, and G. L. Grinblat, and L. Uzal, and S. Rawat, and J. Feist, and L. Mounier, “Toward large-scale vulnerability discovery using machine
learning,” in Proceedings of the Sixth ACM Conference on Data and Application Security and Privacy, 2016, pp. 85–96.

\bibitem{yamaguchi2013chucky}
F. Yamaguchi, and C. Wressnegger, and H. Gascon, and K. Rieck, “Chucky: Exposing missing checks in source code for vulnerability
discovery,” in Proceedings of the 2013 ACM SIGSAC conference on Computer \& communications security, 2013, pp. 499–510.

\bibitem{yamaguchi2012generalized}
F. Yamaguchi, and M. Lottmann, and K. Rieck, “Generalized vulnerability extrapolation using abstract syntax trees,” in Proceedings
of the 28th Annual Computer Security Applications Conference, 2012, pp. 359–368.


\bibitem{li2021sysevr}
 Z. Li, and D. Zou, and S. Xu, and H. Jin, and Y. Zhu, and Z. Chen, “Sysevr: A framework for using deep learning to detect software vulnerabilities,”
IEEE Transactions on Dependable and Secure Computing, 2021.

\bibitem{sedaghatbaf2021automated}
A. Sedaghatbaf, and M. H. Moghadam, and M. Saadatmand, “Automated performance testing based on active deep learning,” arXiv
preprint arXiv:2104.02102, 2021.

\bibitem{li2018vuldeepecker}
Z. Li, and D. Zou, and S. Xu, and X. Ou, and H. Jin, and S. Wang, and Z. Deng, and Y. Zhong, “Vuldeepecker: A deep learning-based system for vulnerability
detection,” arXiv preprint arXiv:1801.01681, 2018.

\bibitem{NVD}
“Nvd: National vulnerability database,” 2021.

\bibitem{SARD}
“Sard: Software assurance reference dataset,” 2021.

\bibitem{llvm}
“Llvm intermediate representation,” 2021.

\bibitem{Goodfellow-et-al-2016}
I. Goodfellow, and Y. Bengio, and A. Courville, Deep Learning, MIT Press, 2016, http://www.deeplearningbook.org.

\bibitem{coimbra2021using}
Coimbra, David, Sofia Reis, Rui Abreu, Corina Păsăreanu, and Hakan Erdogmus. "On using distributed representations of source code for the detection of C security vulnerabilities." arXiv preprint arXiv:2106.01367 (2021).

\bibitem{devlin2018bert}
J. Devlin, M. Chang, K. Lee, K. Toutanova, "Bert: Pre-training of deep bidirectional transformers for language understanding", arXiv preprint arXiv:1810.04805, 2018.

\bibitem{sanh2019disilbert}
V. Sanh, L. Debut, J. Chaumond, T. Wolf, "DistilBERT, a distilled version of BERT: smaller, faster, cheaper and lighter", arXiv preprint arXiv:1910.01108, 2019.


%
%
%
%
%
%
%
%
\end{thebibliography}

\end{document}